\documentclass[aps,twocolumn,floatfix, showpacs,eqsecnum,
superscriptaddress,prb]{revtex4}
\usepackage{color}
\usepackage{graphicx}
\usepackage{amsmath}
\usepackage{amssymb}
\usepackage{bm}

\begin{document}

\title{Phonon-kink scattering effect on the low temperature thermal transport in solids}
\author{J. A. M. van Ostaay}
\affiliation{Instituut-Lorentz, Universiteit Leiden, P.O. Box 9506, 2300 RA
Leiden, The Netherlands}
\author{S. I. Mukhin}
\affiliation{Theoretical physics and quantum technologies department, NUST
MISIS, 119991 Moscow, Russia}
\date{\today}
\pacs{72.10.-d, 72.15.Eb, 66.70.-f, 61.72.Lk, 67.80.-s
E-mail: sergei@lorentz.leidenuniv.nl}

\begin{abstract}
We consider contribution to the phonon scattering, in the temperature range of 1K, by the dislocation kinks pinned in the random stress fields in a crystal. The effect of electron-kink scattering on the thermal transport in the normal metals was considered much earlier \cite{Muk86}.  The phonon thermal transport anomaly at low temperature  was demonstrated by experiments in the deformed (bent) superconducting lead samples \cite{Mez79} and in helium-4 crystals \cite{Mez82, Mez84} and was ascribed to the dislocation dynamics. Previously, we had discussed semi-qualitatively the phonon-kink scattering effects on the thermal conductivity of insulating crystals  in a series of papers \cite{mezmuk, ostmukmez}. In this work it is demonstrated explicitly that exponent of the power low in the temperature dependence of the phonon thermal conductivity depends, due to kinks, on the distribution of the random elastic stresses in the crystal, that pin the kinks motion along the dislocation lines. We found that one of the random matrix distributions of the well known Wigner-Dyson theory is most suitable to fit the lead samples experimental data \cite{Mez79}. We also demonstrate that depending on the distribution function of the oscillation frequencies of the kinks, the power low temperature dependences of the phonon thermal conductivity, in principle, may possess exponents in the range of $2\div 5$.  
\end{abstract}

\maketitle
\section{Introduction}
\begin{figure}[htb]
\centerline{\includegraphics[width=0.4\linewidth]{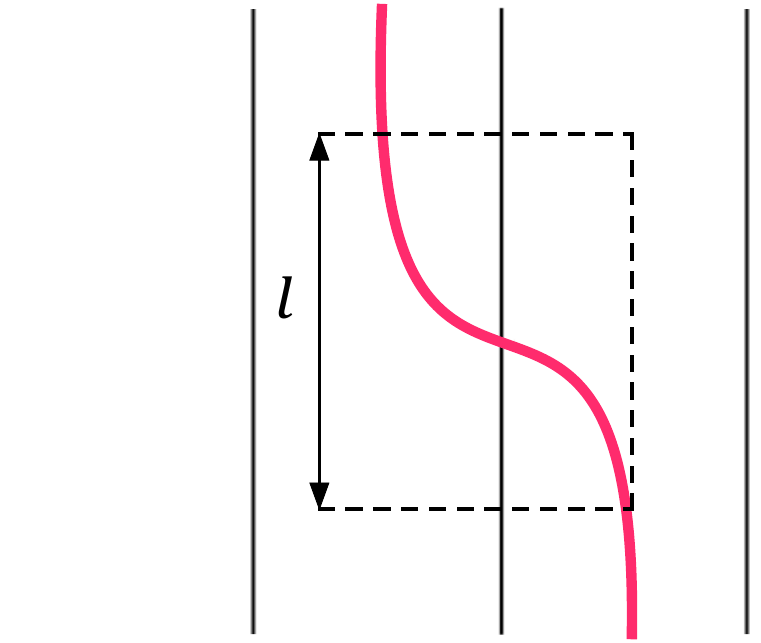}}
\caption{Kink on a dislocation line (red) in the Peierls potential in the crystal (black).}
\label{1}
\end{figure}

It was in the middle of 1983, soon after my PhD thesis defence, that my thesis supervisor Prof. A.A. Abrikosov had introduced me to the head of the Quantum crystals laboratory in Chernogolovka Prof. L.P. Mezhov-Deglin saying: "Sergei, Leonid has a mystery for you to solve". Thus, our collaboration with Leonid Pavlovich has started, and soon evolved into my first publication on the "scattering of electrons by kinks on the dislocation line of a metal" \cite{Muk86}. The major challenge was to find a source of an efficient inelastic spin-conserving scattering of electrons in pure crystals at such a low temperatures ($\leq$ 1K ), that the density of the thermal phonons would be already vanishing. Since the kinks \cite{kinks, kinks1} are topological defects on the dislocation lines in the crystal lattice (Peierls) potential, their density is not vanishing when the temperature goes to zero, unlike the density of the thermal acoustic phonons and/or of the dislocation lines long-wavelength vibrations. The density of kinks, manifesting a topological sector distinct from the ground state of the crystal, depends on the mechanical treatment ('history') of a particular sample.  Hence, e.g. annealing the crystal should remove the low temperature source of inelastic scattering and change the temperature dependence of the thermal conductivity of the same sample when cooling it down again. This idea proposed in my JETP paper \cite{Muk86} was, indeed, in accord with  a particular effect observed by Porf. L.P. Mezhov-Deglin and co-workers, who found that the thermal transport anomalies had disappeared after wearing a sample crystal inside the jacket's pocket for a week or so \cite{Mez79}. Some years after the paper in JETP was published, blown by the wind of 'Perestroika', we met with Prof. L.P. Mezhov-Deglin in the Leiden University, were I served as a postdoc in the group of Prof. Jos de Jongh at the Kammerlingh Onnes Laboratory. Prof. Mezhov-Deglin then drew my attention to the just published paper by the Belgian experimentalists D. Fonteyn and G. Pitsi \cite{font} who had measured torsion dependent heat transport in pure copper single crystals in the He-3 temperatures range and  found  the idea of kinks being a source of low temperature inelastic scattering in solids the most plausible one.  But, only long after the end of 'Perestroika' we had met again with Prof. Mezhov-Deglin and, in that time my Dutch PhD student, Jan van Ostaay at the quiet Chernogolovka premises in the autumn of 2011 and decided to revitalise the investigation of the kink scenario, but now also for the description of the thermal transport anomalies in the crystals with dominating phonon rather than electron thermal flow \cite{Mez82, Mez84}. This ignited our most recent activities \cite{mezmuk, ostmukmez}. The present work is a new logic step in the ongoing research. Namely, we had considered kink-on-dislocation picture in a more fine detail, paying attention to the distribution of the local microscopically 'frozen' stresses in the crystal that provide effective pinning of the kinks motion along the dislocation lines. This effect is described below by an introduction of the distribution function for the kink oscillation frequencies in the random potential of the frozen stress fields in analogy with the introduced long ago by Anderson and co-workers \cite{ahv} distribution of energy splittings of the two level systems in glasses and in spin-glasses \cite{abmu}. As a result, we found that the power low temperature dependences of the phonon thermal conductivity would possess exponents in the range of $2\div 5$ depending on the power law exponent of the frequency-dependent pre-factor in the Wigner-Dyson like distribution function for the kink oscillation frequencies. Another source of randomness related with the kinks comes from the strong anisotropy of the phonon-kink scattering form factor. Namely, a dislocation line breaks translational invariance of the crystal in the plane perpendicular to its axis, while the kink breaks translational invariance along the dislocation axis itself. Correspondingly, the deformation field in the perpendicular to a dislocation axis plane is long ranged and scatters phonons with the wave vectors in the interval $\{0,1/a\}$ ($a$ is  a characteristic radius of the dislocation core). On the other hand, along the dislocation axis only phonons with wave vectors of the order of $~1/l$ are scattered efficiently ($l$ is the kink's length), see Fig.\ref{1}.  When a kink moves along the dislocation axis the deformation field in the perpendicular plane becomes time dependent and causes inelastic scattering of the phonons with the different in-plane wave vectors, provided the conservation laws are obeyed. Simultaneously, phonons with wave vectors $~1/l$  along the dislocation axis can be scattered inelastically by a kink as long as their frequency $\omega\approx s/l$ is close to the kink vibration frequency $\Omega$ ($s$ is an acoustic phonon velocity). Since dislocation lines in a crystal may lay along the different crystal axes, the described above anisotropy in the phonon-kink scattering must be averaged over the different orientation angles of the dislocation lines in a crystal lattice frame. This distribution of the angles depends on the e.g. peculiarities of the deformation process that induces dislocations in a crystal sample \cite {Mez79,Mez84,Muk86}. Another complication, that should be taken into account is the intrinsic renormalisation of the phonon and kink frequencies caused by the phonon-kink interaction via the dislocation line itself. Hence, the mathematical description of the physical phenomenon investigated in this work proves to be straightforward but rather involved. We'll try in the next Sections to avoid as much as possible  the technical details of the bulky analytical derivations in favour of the description of the physically meaningful results. 

The paper is organised as follows. In Section II, following the general method of Ninomiya \cite{Nin68}, we introduce atomic displacement field in a crystal caused by a kink on the edge dislocation line and derive effective mass and bare Hamiltonian of mobile kink. We also derive Hamiltonian of the phonon-kink anisotropic scattering purely from kinematics of the crystal lattice with a dislocation. In Section III kinetic equation for the phonon thermal transport allowing for the phonon-kink scattering is solved and corresponding contribution to the thermal conductivity is calculated. In Section IV different distribution functions for the random kink pinning potential are applied and corresponding different temperature dependences of the thermal conductivity are derived. Theoretically calculated  exponents characterising power law temperature dependences of the thermal conductivity due to phonon-kink scattering are compared with experimental curves and the most relevant version of the distribution function is selected.  In Section V general possibility to 'read' deformation history of pure crystals  by measuring their thermal transport anomalies is discussed.            

\section{Classical Kinematics}
To derive Hamiltonian of the phonon-kink anisotropic scattering based on the kinematics of the crystal lattice with a dislocation we use  the procedure formulated in \cite{Nin68}. For a dislocation line along the axis $z$, carrying the multiple kinks, a displacement of the dislocation core from the straight line in the glide plane, $\xi(z)$,  can be decomposed as follows:
\begin{equation}
\xi(z) = \sum_{n}\sum_\kappa
\xi_0(\kappa)e^{i\kappa(z-z^{n}_0(t))} +
\sum_\kappa\xi(\kappa,t)e^{i\kappa z},
\label{kink}
\end{equation}
where $z^n_0(t)$ is the time-dependent position of the $n$th kink on the dislocation line, $\xi(\kappa,t)$ is the Fourier transform of the shape of the dislocation line and $\xi_0(\kappa)$ is the Fourier transform of the shape of the dislocation line near a kink (see e.g.
\cite{Muk86}):

\begin{equation}
\xi_0(z) = \frac{2a}{\pi}\arctan\left\{\exp\left[\pm\frac{2\pi}{a}(z-z_0(t))\sqrt{
\frac{\alpha}{E_0}}\right]\right\}, \label{kinkshape}
\end{equation}
\noindent where $E_0=T(k_z\approx 0)$ is the line tension that characterises dislocation line bending energy (see (\ref{upot}) below),
$\alpha$ is the height of the crystal lattice Peierls barrier (the dimension is energy per unit of length); $a$ is the period of the valleys
in the Peierls potential, $\nu\equiv \sqrt{\alpha/E_0}\sim 10^{-4}$ (for copper). A corresponding Fourier transform is:

\begin{equation}
\kappa\xi_0(\kappa) \approx \frac{2\pi ia}{L} \exp\left(-\frac{|\kappa
a|}{2}\sqrt{\frac{E_0}{\alpha}}\right).\label{kinkF}
\end{equation}
\noindent The Cartesian components of the lattice displacement vector, $u_j$, around the dislocation line can be thus decomposed into three constituents:
\begin{equation}
u_j = u_j^k + u_j^d + u_j^{ph}, 
\label{displace}
\end{equation}
\noindent where small displacement of the lattice $u_j^d$, induced by dislocation line vibration, and an incident phonon induced lattice displacement $u_j^{ph}$, linearly superimpose with the finite kink induced displacement $u_j^k$. By the virtue of Eq.~\eqref{kink}, the kink contribution $u_j^k$ can be written as:
\begin{equation}
u_j^k = \sum_{n}\sum_\kappa
f_j(\mathbf{r}_{\perp}:\kappa)\xi_0(\kappa)e^{i\kappa(z-z^{n}_0(t))},
\end{equation}
where $f_j(\mathbf{r}_{\perp}:\kappa)$ is a form factor of the dislocation in the crystal lattice that linearly translates a deformation of the dislocation line into deformation of the lattice  around it in the whole crystal. The
abbreviation $\mathbf{r}_{\perp}$ indicates $(x,y)$. Real-valuednes of $u_j^k$ implies
\begin{equation}
\xi^\ast_0(\kappa) = \xi_0(-\kappa), \quad f^\ast_j(\mathbf{r}_{\perp}:\kappa)
= f_j(\mathbf{r}_{\perp}:-\kappa).
\end{equation}
The dislocation line contribution $u^d_j$ equals:
\begin{equation}
u^d_j = \sum_\kappa f_j(\mathbf{r}_\perp : \kappa)\xi(\kappa,t)e^{i\kappa z},
\end{equation}
where real-valuedness of $u^d_j$ implies $\xi^\ast(\kappa,t) = \xi(-\kappa,t)$.
The phonon contribution $u_j^{ph}$ can be expressed as a superposition:
\begin{equation}
u_j^{ph} =
\sum_{\mathbf{k},s}q(\mathbf{k},s)e_j(\mathbf{k},s)e^{i\mathbf{k}\cdot
\mathbf{r}},
\end{equation}
where $s$ indicates the phonon polarisation and $\mathbf{e}$
the polarization vector. Since $u^{ph}_j$ is real, the following identities hold:
\begin{equation}
q^\ast(\mathbf{k},s) = q(-\mathbf{k},s), \quad e^\ast_j(\mathbf{k},s) =
e_j(-\mathbf{k},s).
\end{equation}

The kinetic energy equals:
\begin{equation}
T = \frac{\rho}{2}\int \sum_j (\dot{u}_j^d + \dot{u}_j^k + \dot{u}_j^{ph})^2 dV,
\label{kin}
\end{equation}
where $\rho$ is the (constant) mass density of the crystal and the volume integral runs over the whole sample.  The kinetic energy could be then grouped into four distinct terms:
\begin{equation}
T = T_{q} + T_{d} + T_{k} + T_{\text{mixed}}.
\end{equation}
The kinetic energy solely due to phonons equals:
\begin{equation}
T_{q} = \frac{\rho V}{2}\sum_{\mathbf{k}, s}
\dot{q}(\mathbf{k},s)\dot{q}^\ast(\mathbf{k},s).
\end{equation}
Here, the closure conditions,
\begin{equation}
\sum_s e_j(\mathbf{k},s)e^*_l(\mathbf{k},s) = \delta_{jl}, \label{closure}
\end{equation}
have been used. The kinetic energy for the dislocation line is found to be:
\begin{equation}
T_d = \frac{1}{2}\sum_{k_z} m(k_z)\dot{\xi}(k_z,t)\dot{\xi^\ast}(k_z,t),
\end{equation}
where the mass per mode, $m(k_z)$ is given by
\begin{equation}
m(k_z) = \rho L \sum_{j}\int|f_j(\mathbf{r}_{\perp}:k_z)|^2d^2r_{\perp}.
\label{mass}
\end{equation}
The kinetic energy of the kink is equal to:
\begin{equation}
T_{k} = \sum_{n}\frac{M}{2}(\dot{z}^{n}_0)^2,
\end{equation}
where the kink's mass $M$ equals to:
\begin{equation}
M = \sum_{k_z}k_z^2|\xi_0(k_z)|^2m(k_z). \label{kmass}
\end{equation}
\noindent The smallness of parameter $\nu=\sqrt{\alpha/E_0}\ll 1$ in (\ref{kinkshape}) makes kink a 'light mass' particle: the mass $M$ of the kink is much less than atomic mass in a crystal. This, in turn, provides the smallness (of order $1$ K) of the frequencies of the kink oscillations with respect to the Debye frequency in the crystal (e.g. $\sim 100$ K ) and importance of the inelastic kink scattering for the low temperature $\sim 1$ K heat transport anomalies.
The interaction part that follows directly from the kinematic derivation (\ref{displace})-(\ref{kin}) consists of three terms:
\begin{equation}
T_{\text{mixed}} = T_{d,k} +  T_{q,d} + T_{q,k},
\end{equation}
\noindent where the terms from left to right describe the following interactions: of the kinks with the dislocation vibrations, between the dislocation and the phonons, and between the phonons and the kinks respectively. For the reference purposes we define a complete Fourier transform of the dislocation line form factor $f_j(\mathbf{r}_{\perp}:k_z)$ :

\begin{equation}
F_j(\mathbf{k}) \equiv \frac{1}{L^2}\int e^{-i\mathbf{k}_\perp \cdot
\mathbf{r}_{\perp}} f_j(\mathbf{r}_{\perp}:k_z) d^2r_{\perp}, \label{Fourier}
\end{equation}

\noindent where $\mathbf{k}_\perp = (k_x, k_y)$. With this definition, after somewhat bulky but straightforward algebra one arrives at the following general expression for the total kinetic energy of the crystal lattice vibrations:
\begin{widetext}
\begin{align}
T &= \frac{\rho V }{2} \sum_{\mathbf{k}, s}
\dot{q}(\mathbf{k},s)\dot{q}^\ast(\mathbf{k},s) + \frac{1}{2}\sum_{k_z}
m(k_z)\dot{\xi}(k_z,t)\dot{\xi^\ast}(k_z,t)
+ \sum_n\frac{M}{2}(\dot{Z}^n)^2 + \sum_{\mathbf{k}, s}
\phi_1(\mathbf{k},s)\dot{\xi}^\ast(k_z,t)\dot{q}(\mathbf{k},s)
+\phi^\ast_1(\mathbf{k},s)\dot{\xi}(k_z,t)\dot{q}^\ast(\mathbf{k}, s)\nonumber
\\
&+ i\sum_{n, \mathbf{k}, s}  \left\{(\phi^n_2)^\ast(\mathbf{k},s)
\dot{Z}^n(t) e^{-ik_zZ^n(t)}\dot{q}^\ast(\mathbf{k}, s)
-\phi^n_2(\mathbf{k},s)
\dot{Z}^n(t) e^{ik_zZ^n(t)}\dot{q}(\mathbf{k}, s) \right\} \nonumber \\
&+ \frac{2i}{\rho V}\sum_{n, \mathbf{k}, s}
\left\{\phi_1(\mathbf{k},s)(\phi^n_2)^\ast(\mathbf{k},s)
\dot{Z}^n(t) e^{-ik_zZ^n(t)}\dot{\xi}^\ast(k_z,
t)-\phi_1^\ast(\mathbf{k},s)\phi^n_2(\mathbf{k},s)
\dot{Z}^n(t) e^{ik_zZ^n(t)}\dot{\xi}(k_z, t) \right\},\label{kintotal}
\end{align}
\noindent where $Z^n(t) = z^n_0(t)-z^{0,n}_0$, with $z_0^{0,n}$ the equilibrium position of the kink
$n$. The new functions $\phi_1(\mathbf{k},s)$ and $\phi^n_2(\mathbf{k},s)$ are defined as follows:
\begin{equation}
\phi_1(\mathbf{k},s) = \frac{\rho
V}{2}\sum_j F^\ast_j(\mathbf{k})e_j(\mathbf{k},s), \quad \phi^n_2(\mathbf{k},s)
= k_z\xi^\ast_0(k_z)e^{ik_zz^{0,n}_0} \phi_1(\mathbf{k},s)
\label{phidef}.
\end{equation}
\noindent The only model dependent {\it{ad hoc}} parameters are included in the expression for the potential energy $U$ of the lattice.
The latter is represented with the sum of contributions of the lattice phonon modes (phonon modes $\omega_0(\mathbf{k},s)$ characterise the pure lattice), of the dislocation line vibrations (the latter contains the line tension $T(k_z)$ \cite{Nin68}), and of the kink 1D oscillations (the latter contain oscillation frequency $\Omega$ of the kink in the pinning potential \cite{Muk86}) :  

\begin{equation}
U = \frac{\rho V}{2}\sum_{\mathbf{k},s}\omega_0^2(\mathbf{k},s)
q(\mathbf{k},s)q^*(\mathbf{k},s) + \sum_n\frac{M\Omega^2}{2}(Z^n)^2
+\frac{L}{2}\sum_{k_z} k_z^2T(k_z)\xi(k_z,t)\xi^\ast(k_z,t).\; \label{upot}
\end{equation}
\noindent The normal coordinates are defined as follows:

\begin{subequations}
\begin{align}
Q(\mathbf{k},s)&=q(\mathbf{k},s) + \frac{2\phi^\ast_1(\mathbf{k},s)}{\rho
V}\xi(k_z), \\
Q^\ast(\mathbf{k},s)&=q^\ast(\mathbf{k},s) + \frac{2\phi_1(\mathbf{k},s)}{\rho
V}\xi^\ast(k_z),
\end{align}
\end{subequations}
\noindent and the respective conjugated momenta to the coordinates $Z^n(t)$ and $Q(\mathbf{k},s)$ : $P^n_Z$ and $P_Q(\mathbf{k},s)$, are introduced. 
Then, the total Lagrangian of the crystal can be reconstructed as: $L = T - U$. The Legendre transformation leads to the following Hamiltonian of the crystal lattice:
\begin{equation}
H= H_Q + H_Z + H_{Q,Z}+H_{Q,Q}  \label{Hamiltonian}.
\end{equation}
\noindent Here the different terms in the sum (\ref{Hamiltonian}) are equal to:

\begin{subequations}
\begin{align}
&H_Q = \frac{1}{2}\sum_{\mathbf{k},s}\left\{\frac{(\hat{P}_Q(\mathbf{k},s))^2}{\rho
V}+\rho V\omega^2_0(\mathbf{k},s)(\hat{Q}(\mathbf{k},s))^2\right\} \\
&H_Z = \frac{1}{2}\sum_n\left \{\frac{(\hat{P}_Z^n)^2}{M} +M\Omega^2 (\hat{Z}^n)^2
\right \} \\
&\hat{H}_{Q,Z} = \frac{i}{\rho
V M}\sum_{n,\mathbf{k},s}\left\{\frac{
\phi_2^\ast(\mathbf{k},s)
\hat{P}_Z^n e^{-ik_z\hat{Z}^n}\hat{P}_Q(\mathbf{k}, s)}{(1-\Xi(\mathbf{k},s)\omega^2_0(\mathbf{k},s))}- \frac{\phi_2(\mathbf{k},s)\hat{P}^\dagger_Q(\mathbf{k}, s)e^{ik_z\hat{Z}^n}\hat{P}^{n\dagger}_Z}{(1-\Xi(\mathbf{k},s)\omega^2_0(\mathbf{k},s))}\right\},
\label{interaction_quantum_kink} \\
&\hat{H}_{Q,Q} = \sum_{\mathbf{k},s;\mathbf{k}',s'}\frac{\delta_{k_z,k'_z}
\omega^2_0(\mathbf{k},s)\omega^2_0(\mathbf{k}',s')\left[
\phi_1(\mathbf{k},s)\phi^\ast_1(\mathbf{k}',s')
\hat{Q}(\mathbf{k},s)\hat{Q}^\dagger(\mathbf{k}',s')+\phi^\ast_1(\mathbf{k},
s)\phi_1(\mathbf{k}',s')
\hat{Q}(\mathbf{k}',s')\hat{Q}^\dagger(\mathbf{k},s)\right]}
{(1-\Xi_z(\mathbf{k},s)\omega^2_0(\mathbf{k},s))\left(m(k_z)\Omega^2_{k_z}+ \frac{4}{\rho
V}\sum_{\mathbf{k_\perp},s}|\phi_1(\mathbf{k},s)|^2\omega^2_0(\mathbf{k},s)\right)}
\label{HQZ}
\end{align}
\end{subequations}
\noindent where $\Omega^2_{k_z} \equiv LT(k_z)k^2_z/m(k_z)$, and the resonant denominators in (\ref{interaction_quantum_kink}), (\ref{HQZ}) are:
\begin{equation}
\Xi(\mathbf{k}_0,s_0) = \sum_{\mathbf{k},s}\frac{4|\phi_2(\mathbf{k},s)|^2\omega^2_0(\mathbf{k},s)}{\rho VM\Omega^2(\omega^2_0(\mathbf{k},s)-\omega^2 +i\delta)},\label{resk}
\end{equation}
\noindent and
\begin{equation}
\Xi_z(\mathbf{k}_0,s_0) = \sum_{\mathbf{k}_\perp,s}\frac{4|\phi_1(\mathbf{k},s)|^2\omega^2_0(\mathbf{k},s)}{\rho Vm(k_z)\Omega^2_{k_z}(\omega^2_0(\mathbf{k},s)-\omega^2 +i\delta)},\label{resd}
\end{equation}
\noindent where $\delta=0^+$ is used to regularise the expressions.
Hence, we have derived the phonon-phonon and phonon-kink scattering Hamiltonians in the crystal with the kinks on the dislocation lines, by using only general kinematic approach of Ninomiya \cite{Nin68}, that takes into account the topological nature of these defects. It is straightforward now to formulate kinetic equations for the crystal under a temperature gradient and find contributions to the thermal conductivity from the phonon-dislocation and phonon-kink scattering mechanisms. Certainly, it is no need to say, that in the expressions entering (\ref{Hamiltonian}) all the coordinates and momenta must be understood as the second quantised operators, which is achieved using the following relations \cite{Bel06}:
\begin{subequations}
\begin{align}
\hat{Q}(\mathbf{k},s) &= \sqrt{\frac{\hbar}{2\rho
V\omega_0(\mathbf{k},s)}}\left(\hat{c}^\dagger_{-\mathbf{k},s}+\hat{c}_{\mathbf{
k},s}\right) ,\;\hat{P}_Q(\mathbf{k}, s) =
i\sqrt{\frac{\hbar\rho V\omega_0(\mathbf{k},s)}{2}}
\left(\hat{c}^\dagger_{-\mathbf{k},s}-\hat{c}_{\mathbf{k},s}\right), \\
\hat{Z}^n &= \sqrt{\frac{\hbar}{2M\Omega}}\left(\hat{a}_n^\dagger
+\hat{a}_n\right),\; {\hat{P}^n}_Z = i\sqrt{\frac{\hbar\Omega M}{2}}\left(\hat{a}_n^\dagger -
\hat{a}_n\right).
\end{align}
\label{operators}
\end{subequations}

\noindent  Here it is important to mention how we deal with the sums over the kinks, $\sum_n$, in the equations like (\ref{interaction_quantum_kink}). Namely, we use averaging over the kinks positions in the same fashion as the impurity averaging is done in the Feynman diagrammatic technique of metal alloys, see e.g. \cite{agd}. Hence, probability of the phonon scattering by kinks is proportional to the density of kinks multiplied by a single kink scattering cross section. The interaction Hamiltonian is assumed to be a small perturbation to the lattice Hamiltonian, since the density of dislocations and also the linear density of kinks along the dislocation lines are considered to be small enough. Furthermore, when taking into account the interactions, we will ignore mixing of the terms $\hat{H}_{Q,Z}$ and $\hat{H}_{Q,Q}$, thus disregarding the processes that lead to a finite relaxation time $\tau_k$ of the kink oscillations. The latter is introduced below as a phenomenological constant.

\section{Kinetic equation for the heat transport}
The number of phonons in a state described by the wave vector $\mathbf{k}$ and
polarisation $s$ is indicated by $N_{\mathbf{k}s}$. This number changes due to
interaction of the phonons with the dislocations or with the kinks on the dislocation lines. Therefore
\begin{equation}
\frac{dN_{\mathbf{k}s}}{dt} =
\left(\frac{dN_{\mathbf{k}s}}{dt}\right)_{\text{dislocations}} +
\left(\frac{dN_{\mathbf{k}s}}{dt}\right)_{\text{kinks}}.
\end{equation}
As was mentioned above, we assume that the dislocations and kinks random positions average out, meaning that we can
write the rates of changes as a sum of independent single scattering events:
\begin{equation}
\left(\frac{dN_{\mathbf{k}s}}{dt}\right)_{\text{dislocations}} =
N_d\left(\frac{dN_{\mathbf{k}s}}{dt}\right)_{d}, \quad
\left(\frac{dN_{\mathbf{k}s}}{dt}\right)_{\text{kinks}} =
N_dN_k\left(\frac{dN_{\mathbf{k}s}}{dt}\right)_{k},
\end{equation}
with $N_d$ being the number of dislocations in the crystal, $N_k$ being the
average number of the kinks per dislocation.
For both events, scattering on a dislocation or scattering on a kink, we can
write the scattering rates:
\begin{equation}
\left(\frac{dN_{\mathbf{k}s}}{dt}\right)_j =
\sum_{\mathbf{k}',s'}\left(w_j(\mathbf{k},s;\mathbf{k}',s')-
w_j(\mathbf{k}',s';\mathbf{k},s)\right),\label{rates}
\end{equation}
\noindent where $j=d,k$ mark the dislocation or kink as a scattering source, and the scattering rate $w_j(\mathbf{k},s;\mathbf{k}',s')$ is the probability per unit time for one phonon with wavevector $\mathbf{k}'$ and polarisation $s'$ to scatter into one with wavevector $\mathbf{k}$ and polarisation $s$.

The scattering cross sections could be inferred from the Hamiltonian (\ref{Hamiltonian})-(\ref{HQZ}) and equal: for the scattering on a dislocation: 

\begin{align}
&w_d(\mathbf{k},s;\mathbf{k}',s') = \mathcal{A}_d(\mathbf{k},s;\mathbf{k'},s') N_{\mathbf{k}'s'}(N_{\mathbf{
k}s}+1), \quad \text{with} \label{ad} \\
&\mathcal{A}_d(\mathbf{k},s;\mathbf{k'},s') =
\frac{8\pi\omega'^3_0\omega^3_0|\phi_1(\mathbf{k}',
s')\phi_1(\mathbf{k},s)|^2\delta_{k_z,k'_z}
\delta\left(\omega_0-\omega'_0\right)}
{\left|1-\Xi_z(\mathbf{k},s)\omega^2_0\right|^2\left(\rho Vm(k_z)\Omega^2_{k_z}+4\sum_{\mathbf{k_\perp},s}|\phi_1(\mathbf{k},
s)|^2\omega^2_0\right)^2} \nonumber.
\end{align}
\noindent and for the phonon scattering on a kink:

\begin{align}
\mathcal{A}_k(\mathbf{k},s;\mathbf{k}',s') &=
\frac{\pi^2\tau_k\Omega^2\omega_0\omega'_0|\phi_2(\mathbf{k}',
s')\phi_2(\mathbf{k},s)|^2}{M^2\rho^2 V^2|1-\Xi(\mathbf{k},s)\omega^2_0|^2}[3+ 8N^0(\Omega) + 8(N^0(\Omega))^2]
\delta(\omega_0-\omega'_0)\delta(\omega_0-\Omega), \label{ak}
\end{align}
where we introduced the short-hand notation $\omega_0=\omega_0(\mathbf{k},s)$
and $\omega'_0 = \omega_0(\mathbf{k}',s')$, and wrote the $N^0(\omega)$ for the
Bose-Einstein distribution, $N^0(\omega) = (\exp(\hbar\omega/k_BT)-1)^{-1}$.

Now we are in a position to derive the kinetic equation for the case of a small temperature gradient in a crystal sample with dislocations and kins in order to find phonon based thermal conductivity.  Adding the rates of change of the number of phonons in the crystal due to the different scattering sources we find:

\begin{equation}
\frac{dN_{\mathbf{k}s}}{dt} = N_d\left(\frac{dN_{\mathbf{k}s}}{dt}\right)_{d} + N_dN_k\left(\frac{dN_{\mathbf{k}s}}{dt}\right)_{k},
\label{kinetic_equation_start}
\end{equation}
where for both the kinks and the dislocations we can write:
\begin{equation}
\left(\frac{dN_{\mathbf{k}s}}{dt}\right)_j =
\sum_{\mathbf{k}',s'}\left(w_j(\mathbf{k},s;\mathbf{k}',s')-
w_j(\mathbf{k}',s';\mathbf{k},s)\right),
\end{equation}
\noindent where now we can use (\ref{ad}) and (\ref{ak}) for $w_d(\mathbf{k},s;\mathbf{k}',s')$ and $w_k(\mathbf{k},s;\mathbf{k}',s')$. On the other hand,  in a dynamic equilibrium with a small constant temperature gradient across the sample the time derivative of $N_{\mathbf{k}s}$ has to be read as:
\begin{equation}
\frac{dN_{\mathbf{k}s}}{dt} = \frac{\partial N_{\mathbf{k}s}}{\partial
\mathbf{r}} \cdot \dot{\mathbf{r}} \approx \frac{\partial N^0}{\partial T}\mathbf{\nabla}T \cdot \frac{\partial
\omega_0(\mathbf{k},s)}{\partial \mathbf{k}} =
\frac{\hbar\omega_0(\mathbf{k},s)}{k_BT^2}
N^0(\omega_0(\mathbf{k},s))(1+N^0(\omega_0(\mathbf{k},s))\nabla T \cdot
\frac{\partial \omega_0({\mathbf k},s)}{\partial {\mathbf k}},
\end{equation}
\noindent where $\partial \omega_0(\mathbf{k},s)/\partial \mathbf{k}$ is the phonon
velocity. Here the phonon distribution function $N_{\mathbf{k}s}$ is substituted with its unperturbed value of the Bose-Einstein distribution, $N^0(\omega_0(\mathbf{k},s))$ in the linear approximation with respect to the temperature gradient and to the small deviation $\delta N_{\mathbf{k}s}\sim \nabla T$ defined as:  

\begin{equation}
N_{\mathbf{k}s}(\omega_0({\mathbf k},s)) = N^0 (\omega_0(\mathbf{k},s)) + \delta
N_{\mathbf{k}s}.
\end{equation}
These small fluctuations typically do not contribute to the spatial derivative
of the phonon distribution function.When there is a temperature gradient present in the system, that will cause it. 
For the purpose of the following derivation it is useful to prove that:
\begin{equation}
w_j(\mathbf{k}',s';\mathbf{k},s)
=w_j(\mathbf{k},s;\mathbf{k}',s')\exp\{\hbar(\omega_0({{\mathbf k}},s)-
\omega_0({{\mathbf k'}},s'))/k_BT\}
\frac{N_{\mathbf{k}s}(\omega_0(\mathbf{k},s))(1+N_{\mathbf{k}'s'}
(\omega_0(\mathbf{k}',s')))}{N_{\mathbf{k}s}(\omega_0(\mathbf{k}',s'))(1+N_{
\mathbf{k}s}(\omega_0(\mathbf{k},s)))} \label{equality}.
\end{equation}
The proof of this goes as follows. In $w_j(\mathbf{k}',s'; \mathbf{k}, s)$ the
reverse process with respect to $w_j(\mathbf{k},s;\mathbf{k}',s')$ is considered, therefore, the $(\mathbf{k},s)$ and
$(\mathbf{k}',s')$ states have to be interchanged. For the kink-part this
interchange implies that now $E$ plays the role of the original energy, while
$E'$ plays the role of final energy. As $w$ includes energy conservation:
$\delta(\hbar(\omega_0(\mathbf{k},s)-\omega_0(\mathbf{k'},s'))+(E-E'))$, the thermal
averaging for $w_j(\mathbf{k}',s'; \mathbf{k}, s)$ can be written as:
\begin{equation}
\sum_{E, E'} e^{-(E'+f)/k_BT} =
\sum_{E,E'}e^{-(E-\hbar(\omega_0(\mathbf{k},s)-\omega_0(\mathbf{k}',s'))+f)/k_B
T} = e^{\hbar(\omega_0(\mathbf{k},s)-\omega_0(\mathbf{k}',s')/k_BT}\sum_{E,E'}
e^{-(E+f)/k_BT},
\end{equation}
\noindent where $f(E,E')$ is some function.
As the thermal averaging demands that the initial energy is put in the exponential,
the last term in the expression above has to be used in $w_j(\mathbf{k}',s';
\mathbf{k}, s)$, thus, explaining the exponential arising in Eq.~\eqref{equality}.

Using this equality we end up with a kinetic equation that we are going to use in the next Section:
\begin{align}
&\frac{\hbar\omega_0(\mathbf{k},s)}{k_BT^2} N^0(\omega_0(\mathbf{k},s))(1+N^0(\omega_0(\mathbf{k},s)) \nabla T \cdot
\frac{\partial \omega_0({\mathbf k},s)}{\partial {\mathbf k}} =
N_dV\sum_{s'}\int \frac{d^3 k'}{(2\pi)^3} \left(\mathcal{A}_d({\mathbf
k},s;{\mathbf k}',s') + N_k\mathcal{A}_k({\mathbf k},s;{\mathbf k}',s') \right) \nonumber\\
&\times \left\{ N_{{\mathbf k}'s'}(\omega_0({\mathbf k}',s')) (1+N_{\mathbf{
k}s}(\omega_0({\mathbf k},s))) - \exp\{\hbar(\omega_0({{\mathbf k}},s)-
\omega_0({{\mathbf k'}},s'))/k_BT\} N_{\mathbf{k}s} (\omega_0({\mathbf
k},s))(1+N_{{\mathbf
k}'s'}(\omega_0({\mathbf k}',s')) \right\},
\label{kinetic_equation}
\end{align}

\section{Thermal conductivity}
The calculation of the thermal conductivity will be done in analogy with \cite{Muk86}. We start with the kinetic equation, that is derived in the previous Section.  The right-hand-side of Eq.~\eqref{kinetic_equation} is strictly zero for the
Bose-Einstein distribution. Therefore, keeping only terms linear in $ \delta N_{\mathbf{k}s}$ we find:
\begin{align}
&\frac{\hbar\omega_0(\mathbf{k},s)}{k_BT^2}
N^0(\omega_0(\mathbf{k},s))(1+N^0(\omega_0(\mathbf{k},s)) \nabla T \cdot
\frac{\partial \omega_0({\mathbf k},s)}{\partial {\mathbf k}} = \nonumber \\
&N_dV\sum_{s'}\int \frac{d^3 k'}{(2\pi)^3} \left(\mathcal{A}_d({\mathbf
k},s;{\mathbf k}',s') + N_k\mathcal{A}_k({\mathbf k},s;{\mathbf k}',s') \right) \left\{ -\delta N_{\mathbf{k}s}
\frac{N^0(\omega_0(\mathbf{k}',s'))}{N^0(\omega_0(\mathbf{k},s))}
+ \delta N_{\mathbf{k}'s'} \frac{N^0(\omega_0(\mathbf{k},s)) +
1}{N^0(\omega_0(\mathbf{k}',s')) + 1}  \right\}.
\end{align}
This can be rewritten as
\begin{equation}
-\frac{\hbar\omega_0(\mathbf{k},s)}{k_BT^2}  ~ N^0(\omega_0(\mathbf{k},s))(1+N^0(\omega_0(\mathbf{k},s))~\nabla T \cdot \frac{\partial \omega_0({\mathbf k},s)}{\partial {\mathbf k}} =
\sum_{s'}\int \frac{d^3 k'}{(2\pi)^3} \mathcal{P} ({\mathbf k},s;{\mathbf k}',s') [\delta \tilde{N}_{\mathbf{k}s} - \delta \tilde{N}_{\mathbf{k}'s'} ], \label{kinetic_equation_final}
\end{equation}
with
\begin{equation}
\delta \tilde{N}_{\mathbf{k}s} = \frac{\delta N_{\mathbf{k}s}}{N^0(\omega_0(\mathbf{k},s))(1+N^0(\omega_0(\mathbf{k},s)))}.
\end{equation}
and
\begin{equation}
\mathcal{P} ({\mathbf k},s;{\mathbf k}',s') = N_dV \left(\mathcal{A}_d({\mathbf
k},s;{\mathbf k}',s') + N_k\mathcal{A}_k({\mathbf k},s;{\mathbf k}',s') \right) N^0(\omega_0(\mathbf{k}',s'))(1+N^0(\omega_0(\mathbf{k},s))).\label{scatmat}
\end{equation}

The heat flow $\mathbf{Q}$ is given by
\begin{equation}
\mathbf{Q} = \sum_s\int \frac{d^3k}{(2\pi)^3} \hbar \omega_0(\mathbf{k},s)
\frac{\partial \omega_0(\mathbf{k},s)}{\partial \mathbf{k}} \delta
N_{\mathbf{k}s}  \approx
-\hat{\chi} \nabla T\label{heat_flow},
\end{equation}
where $\hat{\chi}$ is the matrix of the thermal conductivity tensor. For simplicity, we will
assume that this matrix only has two distinct diagonal elements and no
off-diagonal elements:
\begin{equation}
\hat{\chi} = \begin{pmatrix}
\chi_\perp & 0 & 0 \\
0 & \chi_\perp & 0 \\
0 & 0 & \chi_\parallel
\end{pmatrix}.
\end{equation}
This implies that there are two distinct heat flows: one along the dislocation line:
\begin{equation}
Q_\parallel = -\chi_\parallel (\nabla T)_z,
\end{equation}
and one perpendicular to it:
\begin{equation}
\mathbf{Q}_\perp = - \chi_\perp (\nabla T)_\perp,
\end{equation}
with $(\nabla T)_\perp = ((\nabla T)_x, (\nabla T)_y, 0)$.

By multiplying both sides of Eq.~\eqref{kinetic_equation_final} with
$\sum_s\int\frac{d^3k}{(2\pi)^3}\delta\tilde{N}_{\mathbf{k}s}$, one finds
\begin{equation}
(\mathbf{\nabla}T) \cdot \mathbf{Q} = -k_BT^2 \sum_{s,s'}\int
\int \frac{d^3k}{(2\pi)^3}\frac{d^3k'}{(2\pi)^3}\delta
\tilde{N}_{\mathbf{k}s}\mathcal{P} ({\mathbf k},s;{\mathbf
k}',s')\left[\delta\tilde{N}_{\mathbf{k}s} - \delta\tilde{N}_{\mathbf{k}'s'}
\right].
\end{equation}
Due to the double integral over both $\mathbf{k}$ and $\mathbf{k}'$ in the expression above, it can also be written as
\begin{equation}
(\mathbf{\nabla}T) \cdot \mathbf{Q} =
-\frac{k_BT^2}{2} \sum_{s,s'}\int \int
\frac{d^3k}{(2\pi)^3}\frac{d^3k'}{(2\pi)^3}\left[\delta\tilde{N}_{\mathbf{k}s} -
\delta\tilde{N}_{\mathbf{k}'s'} \right]\mathcal{P} ({\mathbf k},s;{\mathbf
k}',s')\left[\delta\tilde{N}_{\mathbf{k}s} - \delta\tilde{N}_{\mathbf{k}'s'}
\right].
\end{equation}
Combining this result with equation Eq.~\eqref{heat_flow}, we find the general expression that we'll use to calculate the thermal conductivity:
\begin{equation}
\chi_j^{-1} =
\frac{k_B T^2}{2}\sum_{s,s'}\int \int d^3kd^3k'\mathcal{P} ({\mathbf
k},s;{\mathbf k}',s')\left[\delta\tilde{N}_{\mathbf{k}s}
- \delta\tilde{N}_{\mathbf{k}'s'} \right]^2
\left(\sum_s\int d^3k \hbar \omega_0(\mathbf{k},s) \left(\frac{\partial
\omega_0(\mathbf{k},s)}{\partial \mathbf{k}}\right)_z \delta N_{\mathbf{k}s}
\right)^{-2}, \label{chi_j}
\end{equation}
\noindent where $j=_\parallel,_\perp$ for a particularly oriented dislocation axis.

Using a variational approach~\cite{Zim60}, we write $\delta N_{\mathbf{k}s}$ and $\delta \tilde{N}_{\mathbf{k}s}$  as
\begin{equation}
\delta N_{\mathbf{k}s} = -\left[\frac{\partial N^0
(\omega_0(\mathbf{k},s))}{\partial
\omega_0(\mathbf{k},s)}\right]\Psi_{\mathbf{k}s},  \quad
\delta \tilde{N}_{\mathbf{k}s} =  \frac{\delta
N_{\mathbf{k}s}}{N^0(\omega_0(\mathbf{k},s))(1+N^0(\omega_0(\mathbf{k},s)))}
= \frac{\hbar}{k_BT}\Psi_{\mathbf{k}s},
\end{equation}
where $\Psi_{\mathbf{k}s}$ is a trial wave function. For $\chi_\parallel$, we
choose $\Psi_{\mathbf{k}s} \sim
\omega_0(\mathbf{k},s)k_z$ and for $\chi_\perp$, we choose
$\Psi_{\mathbf{k}s} \sim \omega_0(\mathbf{k},s)k_x$.  The denominator $\mathcal{D}_j$ for both cases is given by:
\begin{equation}
\mathcal{D}_j = \left(\sum_s\int d^3k \hbar \omega_0\left(\frac{\partial
\omega_0}{\partial \mathbf{k}}\right)_j \frac{\partial
N^0(\omega_0)}{\partial\omega_0} \omega_0k_j \right)^2.
\end{equation}
\noindent We use the short-hand notation $\omega_0 = \omega_0(\mathbf{k},s)$
and $\omega_0' = \omega_0(\mathbf{k}',s')$.

Using spherical coordinates and repressing the $\mathbf{k}$ and $s$ dependence
of $\omega_0$, one finds
\begin{equation}
\mathcal{D}_\parallel = \left(\sum_s\frac{\hbar^2c_s}{k_BT}\int dk d\vartheta
d\varphi
\omega^2_0
k^3 \cos^2\vartheta\sin\vartheta
\frac{\exp(\hbar\omega_0/k_BT)}{\left(\exp(\hbar\omega_0/k_BT)-1\right)^2}
\right)^2,
\end{equation}
and
\begin{equation}
\mathcal{D}_\perp = \left(\sum_s\frac{\hbar^2c_s}{k_BT}\int dk d\vartheta
d\varphi
\omega^2_0 k^3 \cos^2\varphi\sin^3\vartheta
\frac{\exp(\hbar\omega_0/k_BT)}{\left(\exp(\hbar\omega_0/k_BT)-1\right)^2}
\right)^2.
\end{equation}

We used for the dispersion relation, $\omega_0(\mathbf{k},s) =
c_s||\mathbf{k}||$, where $c_s$ is the speed of sound in the crystal. At the
temperatures that are much smaller than $\hbar\omega_D/k_B$, with $\omega_D$ the
Debye frequency, we then find:
\begin{subequations}
\begin{align}
\mathcal{D}_\parallel &= 80^2\zeta^2(5)\pi^2\hbar^2\left(\sum_s \frac{1}{c^3_s}
\right)^2\left(\frac{k_BT}{\hbar}\right)^{10}, \\
\mathcal{D}_\perp &= 160^2\zeta^2(5)\pi^2\hbar^2\left(\sum_s \frac{1}{c^3_s}
\right)^2\left(\frac{k_BT}{\hbar}\right)^{10},
\end{align}
\end{subequations}
where $\zeta(s) = \sum_{n=1}^\infty \frac{1}{n^s}$ is the Riemann zeta
function. Numerically, $\zeta(5) = 1.03692$.
The numerator $\mathcal{N}_j$ of Eq.~\eqref{chi_j} can be written as:
\begin{equation}
\mathcal{N}_j = \mathcal{N}_{j,k} + \mathcal{N}_{j,d},
\end{equation}
where
\begin{equation}
\mathcal{N}_{j,k} =\frac{\hbar^2V}{2k_B}N_dN_k\sum_{s,s'}\int \int
d^3kd^3k' \mathcal{A}_k(\mathbf{k},s;\mathbf{k}',s')
N^0(\omega_0')(1+N^0(\omega_0))(\omega_0k_j-\omega'_0k'_j)^2,
\end{equation}
and
\begin{equation}
\mathcal{N}_{j,d} =\frac{\hbar^2V}{2k_B}N_d\sum_{s,s'}\int \int
d^3kd^3k' \mathcal{A}_d(\mathbf{k},s;\mathbf{k}',s')
N^0(\omega_0')(1+N^0(\omega_0))(\omega_0k_j-\omega'_0k'_j)^2, \label{N_d}
\end{equation}
In both equations it holds that for $j=\perp$ $k_j=k_x$ and for $j=\parallel$ $k_j = k_z$.
Since we use isotropic dispersion for simplicity: $\omega_0 = c_s||\mathbf{k}||$, we can switch to spherical coordinates in
the following way:
\begin{equation}
\int_{\mathbb{R}^3}d^3k =
\frac{1}{c^3_s}\int^{2\pi}_0\int^\pi_0\int^\infty_0\omega^2_0\sin\vartheta d\omega_0d\vartheta d\varphi.
\end{equation}

\subsection{Heat conductance: phonon-dislocation scattering}

Plugging in the expression for $\mathcal{A}_d(\mathbf{k},s;\mathbf{k}',s')$ and using
spherical coordinates, after tedious but straightforward calculations we arrive at the following final result for the phonon-dislocation scattering:
\begin{align}
\mathcal{N}_{\parallel,d}& =0,\\
\frac{\mathcal{N}_{\perp,d}}{\mathcal{D}_\perp}
&=\frac{n_dL}{1600\zeta^2(5)k^4_Dk_B}\frac{c^4_tc^4_\ell}{\left(\sum_s\frac{1}{c^3_s}\right)^2}
\Bigg(\int^\infty_0d\tilde{\omega}_0\frac{\tilde{\omega}^9_0\exp\left[\tilde{\omega}_0\right]}{\left(\exp\left[\tilde{\omega}_0\right]-1\right)^2}(1-u^2)^2 \nonumber \\
&\times\Big[2\int^1_0du
(1-u^2)^3\left(\frac{1}{c^{11}_t}  + \frac{1}{c^{11}_\ell}\right)\left\{\frac{1}{\tilde{g}_z(\tilde{\omega}_0,u,t)}+\frac{1}{\tilde{g}_z(\tilde{\omega}_0,u,\ell)}\right\}+\int^{\text{min}(1,\frac{c_t}{c_\ell})}_0\frac{du}{c^6_\ell c^{11}_t}(c^2_t-c^2_\ell u)^2\frac{\left\{c^2_t+c^2_\ell-2c^2_\ell u^2\right\}}{\tilde{g}_z(\tilde{\omega}_0,u,t)}\nonumber \\
&+\int^{\text{min}(1,\frac{c_\ell}{c_t})}_0\frac{du}{c^{11}_\ell c^6_t}(c^2_\ell- c^2_t u^2)^2\frac{\left\{c^2_t+c^2_\ell-2c^2_t u^2\right\}}{\tilde{g}_z(\tilde{\omega}_0,u,\ell)} \Big]\Bigg) \label{thermal_d},
\end{align}
\noindent where $u = \cos\vartheta$, $n_d = N_d/L^2$ is the dislocation density, $\tilde{\omega}_0 = \hbar\omega_0/k_BT$ and:
\begin{align}
g_z(\tilde{\omega}_0,u,s) &= \frac{(c^2_t-c^2_\ell)^2}{4c^4_t} + \frac{(c^2_t-c^2_\ell)c^2_\ell}{4c^2_t\ln\left(\frac{\tilde{\omega}^2_D}{\tilde{\omega}^2_0u^2}\right)}
\left[\frac{c^2_s-u^2c^2_t}{u^2c^4_t}\ln\left|\frac{\tilde{\omega}^2_D - \tilde{\omega}^2_0c^2_s/c^2_t}{\tilde{\omega}^2_0u^2-\tilde{\omega}^2_0c^2_s/c^2_t}\right| + \frac{c^2_s-u^2c^2_\ell}{u^2c^4_\ell}\ln\left|\frac{\tilde{\omega}^2_D - \tilde{\omega}^2_0c^2_s/c^2_\ell}{\tilde{\omega}^2_0u^2-\tilde{\omega}^2_0c^2_s/c^2_\ell}\right|\right] \nonumber \\
&+ \frac{c^4_\ell}{4\ln^2\left(\frac{\tilde{\omega}^2_D}{\tilde{\omega}^2_0u^2}\right)}\Bigg[\left(\frac{c^2_s-u^2c^2_t}{u^2c^4_t}\ln\left|\frac{\tilde{\omega}^2_D - \tilde{\omega}^2_0c^2_s/c^2_t}{\tilde{\omega}^2_0u^2-\tilde{\omega}^2_0c^2_s/c^2_t}\right| + \frac{c^2_s-u^2c^2_\ell}{u^2c^4_\ell}\ln\left|\frac{\tilde{\omega}^2_D - \tilde{\omega}^2_0c^2_s/c^2_\ell}{\tilde{\omega}^2_0u^2-\tilde{\omega}^2_0c^2_s/c^2_\ell}\right|\right)^2\nonumber \\
&+4\pi^2\left(\frac{c^2_s-u^2c^2_t}{u^2c^4_t}\theta\left(1-\frac{uc_t}{c_s}\right) + \frac{c^2_s-u^2c^2_\ell}{u^2c^4_\ell}\theta\left(1-\frac{uc_\ell}{c_s}\right) \right)^2\Bigg].
\end{align}
\noindent Here we also defined: $\tilde{\omega}_D = \hbar c_sk_D/k_BT = T_D/T$, and $T_D$ is the Debye temperature. Now, taking into account that for lead:
\begin{equation}
\alpha = \frac{c^2_t}{c^2_\ell(1+(c_t/c_\ell)^4)}\approx  0.106,
\end{equation}
\noindent and averaging over the orientations of the dislocation lines with respect to the temperature gradient:  $\chi^{-1}_d =
\gamma\chi^{-1}_{\parallel,d} + (1-\gamma)\chi^{-1}_{\perp,d},$ with $\gamma
\in [0,1]$, we perform the integrals numerically. The result is shown in the graph Fig.~\ref{dislocationplot}, where temperature dependent contribution to the heat conductance due to the phonon-dislocation scattering is plotted. 

\begin{figure}[htb]
\centerline{\includegraphics[width=0.4\linewidth]{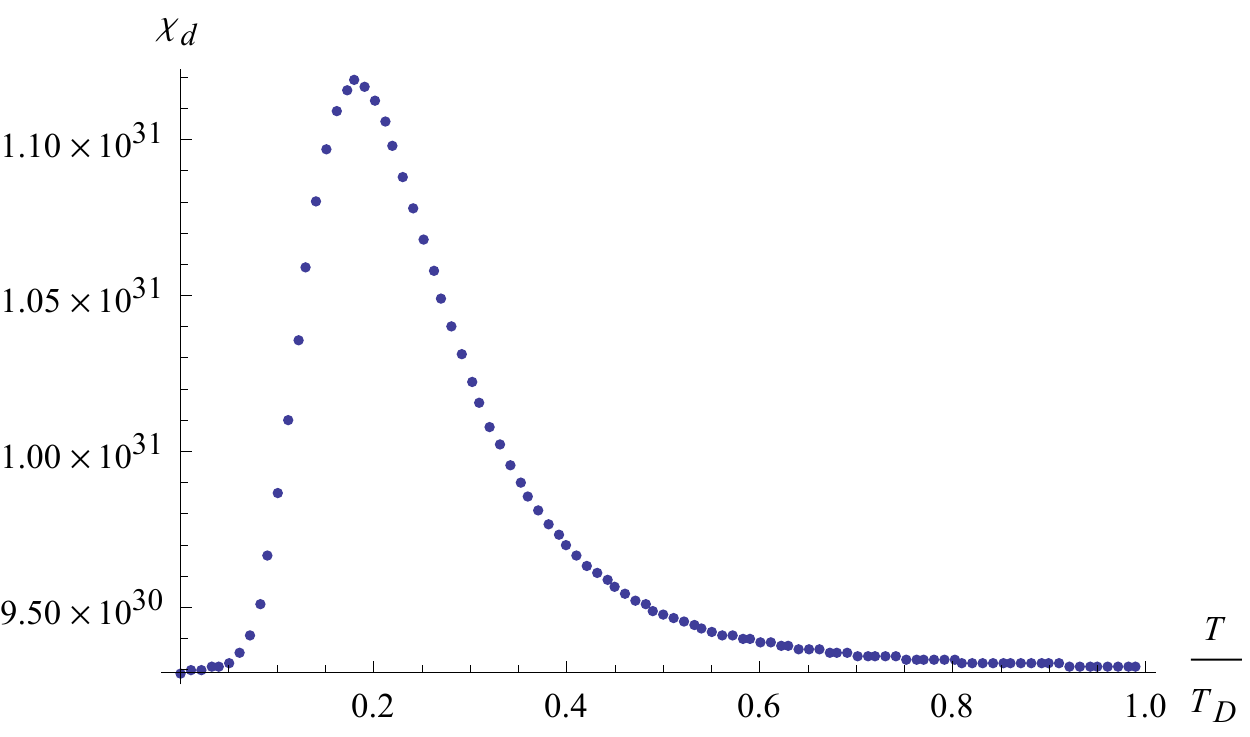}
}
\caption{Temperature dependence (in units of the Debye temperature) of the thermal conductivity (arbitrary units) due to phonon dislocation scattering.}
\label{dislocationplot}
\end{figure}

\subsection{Heat conductance: phonon-kinks scattering}

Now, allowing for the energy conservation in the process: absorbed phonon-excited kink- emitted phonon, as expressed by the delta functions present in Eq.~\eqref{ak}, we find for the phonon-kink scattering the following expressions:
\begin{equation}
\mathcal{N}_{\parallel, k}
= \frac{32\hbar^2\pi^8\tau_k(1+(c_t/c_\ell)^8)\Omega^4\alpha}{
a^2Vk_BE_0}
\frac{N_dN_k\exp[\hbar\Omega/k_BT]}{g(\Omega)\left(\exp[
\hbar\Omega/k_BT]-1\right)^2}\left[3+\frac{8\exp[\hbar\Omega/k_BT]}{\left(\exp[
\hbar\Omega/k_BT]-1\right)^2}\right],
\end{equation}
and
\begin{equation}
\mathcal{N}_{\perp, k} =
\frac{4\hbar^2\pi^8\tau_k\left(1 + (c_t/c_\ell)^4 + (c_t/c_\ell)^6 + (c_t/c_\ell)^{10}\right)\Omega^6}{Vc_t^2
k_B}\frac{N_dN_k\exp[\hbar\Omega/k_BT]}{g(\Omega)\left(\exp[
\hbar\Omega/k_BT]-1\right)^2}\left[3+\frac{8\exp[\hbar\Omega/k_BT]}{\left(\exp[
\hbar\Omega/k_BT]-1\right)^2}\right],
\end{equation}
where
\begin{equation}
g(\Omega) = \left(\frac{2M}{\rho b^2a}\sqrt{\frac{E_0}{\alpha}}- \left(1+\left(\frac{c_t}{c_\ell}\right)^4\right)\ln\frac{k^2_BT^2_D}{\hbar^2\Omega^2}\right)^2 + \pi^2\left(1+\left(\frac{c_t}{c_\ell}\right)^4\right)^2 .
\end{equation}

\noindent Here we assumed that $c_tk_D, c_\ell k_D \gg \Omega$, and that typically $\frac{a\Omega}{c_t}\sqrt{\frac{E_0}{\alpha}} \gg 1$ and $\frac{a\Omega}{c_\ell}\sqrt{\frac{E_0}{\alpha}} \gg 1$~\cite{Muk86}. 
In reality, as we outlined in the Introduction, all the kinks have a slightly different vibration frequencies due to local stress
variations. Therefore, we cannot use a fixed $\Omega$ for the kink and instead have to take a probability distribution for the different frequencies into account. We
will use normalised probabilities based on random matrix theory~\cite{Rab04},
\begin{equation}
P(\Omega) = \frac{b_\beta}{\Delta}
\left(\frac{\Omega}{\Delta}\right)^\beta\exp\left[
-\alpha_\beta\left(\frac{\Omega}{\Delta}\right)^2\right],
\end{equation}
where $\hbar\Delta/k_B \sim 1K$\cite{Muk86}, $\beta=1,2,4$ and~\cite{Bee97} $b_1 = \pi/2$, $a_1 = \pi/4$,
$b_2=32/\pi^2\approx 3.24$, $a_2 = 4\pi$, $b_4=262144/729\pi^3 \approx 11.6$ and
$a_4 = 64/9\pi \approx 2.26$. The $\Omega$ averaged numerator $\tilde{\mathcal{N}}$ is
therefore:
\begin{equation}
\tilde{\mathcal{N}} = \int^\infty_0 d\Omega \mathcal{N}(\Omega)P(\Omega).
\end{equation}
We then find, switching to the variable
$\tilde{\Omega} = \frac{\hbar\Omega}{k_B T}$, the following expressions:
\begin{equation}
\tilde{\mathcal{N}}_{\parallel, k} =
\frac{32b_\beta N_dN_k\pi^8\hbar^2\tau_k(1+(c_t/c_\ell)^8)\alpha}{
a^2Vk_B\Delta^{\beta+1}E_0}\left(\frac{k_B
T}{\hbar}\right)^{\beta+5} \int^\infty_{0} d\tilde{\Omega}
\tilde{\Omega}^{\beta+4}\frac{e^{\tilde{\Omega}-\delta_\beta\tilde{\Omega}^2}}{
g(\tilde{\Omega})\left(e^{\tilde{\Omega}} -1\right)^2}\left[3+\frac {8e^{\tilde{\Omega}}}{\left(e^{\tilde{\Omega}}-1\right)^2}\right],
\end{equation}
and
\begin{equation}
\tilde{\mathcal{N}}_{\perp, k} =
\frac{4 b_\beta\pi^8 N_dN_k \hbar^2\tau_k\left(1 + (c_t/c_\ell)^4 + (c_t/c_\ell)^6 + (c_t/c_\ell)^{10}\right)}{Vc_t^2 k_B\Delta^{\beta+1}} \left(\frac{k_B T}{\hbar}\right)^{\beta
+7}\int^\infty_{0} d\tilde{\Omega}
\tilde{\Omega}^{\beta+6}\frac{e^{\tilde{\Omega}-\delta_\beta\tilde{\Omega}^2}}{
g(\tilde{\Omega})\left(e^{\tilde{\Omega}} -1\right)^2}\left[3+\frac
{ 8e^{\tilde{\Omega}}}{\left(e^{\tilde{\Omega}}-1\right)^2}\right],
\end{equation}
\noindent where
\begin{equation}
\delta_\beta = a_\beta\frac{k^2_BT^2}{\hbar^2\Delta^2}.
\end{equation}
As $\chi_j = \mathcal{D}_j/\tilde{\mathcal{N}}_j$, we thus find
\begin{equation}
\chi^{-1}_{\parallel,k} =
\frac{b_\beta \pi^6 N_dN_k\tau_k(1+(c_t/c_\ell)^8)\alpha}{200\zeta^2(5)a^2
Vk_B\Delta^6E_0\left(\sum_s\frac{1}{c^3_s}
\right)^2}\left(\frac { k_B
T}{\hbar\Delta}\right)^{\beta
-5} \int^\infty_{0} d\tilde{\Omega}
\tilde{\Omega}^{\beta+4}\frac{e^{\tilde{\Omega}-\delta_\beta\tilde{\Omega}^2}}{
g(\tilde{\Omega})\left(e^{\tilde{\Omega}} -1\right)^2}\left[3+\frac {
8e^{\tilde{\Omega}}}{\left(e^{\tilde{\Omega}}-1\right)^2}\right] ,
\end{equation}
and
\begin{equation}
\chi^{-1}_{\perp,k} =
\frac{b_\beta \pi^6 N_dN_k\tau_k\left(1 + (c_t/c_\ell)^4 + (c_t/c_\ell)^6 + (c_t/c_\ell)^{10}\right)}{6400\zeta^2(5)Vc_t^2 k_B\Delta^4\left(\sum_s\frac{1}{c^3_s}
\right)^2}\left(\frac{k_B
T}{\hbar\Delta}\right)^{\beta
-3}\int^\infty_{0} d\tilde{\Omega}
\tilde{\Omega}^{\beta+6}\frac{e^{\tilde{\Omega}-\delta_\beta\tilde{\Omega}^2}}{
g(\tilde{\Omega})\left(e^{\tilde{\Omega}} -1\right)^2}\left[3+\frac
{ 8e^{\tilde{\Omega}}}{\left(e^{\tilde{\Omega}}-1\right)^2}\right].
\end{equation}
In reality no sample is dominated by either only purely parallel or
perpendicular scattering. Therefore we need to average over both parallel and
perpendicular scattering. As conductances add inversely, this implies that the
total conductance $\chi_k$ is found from $\chi^{-1}_k =
\gamma\chi^{-1}_{\parallel,k} + (1-\gamma)\chi^{-1}_{\perp,k},$ with $\gamma
\in [0,1]$. Therefore,
\begin{align}
\chi_k &=
\frac{200\zeta^2(5)k_B\Delta^4E_0\left(\sum_s\frac{1}{c^3_s}
\right)^2}{b_\beta\pi^6n_dn_k\tau_k} \nonumber \\
&\times\Bigg[\gamma\frac{\left(1+(c_t/c_\ell)^8\right)\alpha}{a^2\Delta^2E_0}
\left(\frac {k_BT}{\hbar\Delta}\right)^{\beta-5}
\int^\infty_{0}d\tilde{\Omega}
\tilde{\Omega}^{\beta+4}\frac{e^{\tilde{\Omega}-\delta_\beta\tilde{\Omega}^2}}{
g(\tilde{\Omega})\left(e^{\tilde{\Omega}} -1\right)^2}\left[3+\frac {
8e^{\tilde{\Omega}}}{\left(e^{\tilde{\Omega}}-1\right)^2}\right]\nonumber \\
&+(1-\gamma)\frac{1 + (c_t/c_\ell)^4 + (c_t/c_\ell)^6 + (c_t/c_\ell)^{10}}{32c^2_t}
\left(\frac{k_BT}{\hbar\Delta}\right)^{\beta-3}\int^\infty_{0} d\tilde{\Omega}
\tilde{\Omega}^{\beta+6}\frac{e^{\tilde{\Omega}-\delta_\beta\tilde{\Omega}^2}}{
g(\tilde{\Omega})\left(e^{\tilde{\Omega}} -1\right)^2}\left[3+\frac
{ 8e^{\tilde{\Omega}}}{\left(e^{\tilde{\Omega}}-1\right)^2}\right]\Bigg]^{-1},
\label{Kinks}
\end{align}
where $n_k = N_k/L$ is the linear kink density.  Plugging in realistic values allows us to perform a numerical evaluation of eq.~\eqref{Kinks}. 
Regardless of the value of $\beta$, $\chi_k$ will scale as
$T^6$ for $k_BT/\hbar\Delta \gg 1$. For $k_BT/\hbar\Delta \ll 1$ the thermal conductivity will scale as $T^{2.0}$ for  $\beta=1$, as $T^{3.9}$ for  $\beta=2$, and as $T^{4.8}$ for  $\beta=4$. Therefore, the value $\beta =2$ coincides the most with the experiments~\cite{Mez79}, see Fig.~\ref{kinkplot}, where the different, corresponding to $\beta=1,2,4$, temperature dependencies of the thermal conductivity due to the phonon-kinks scattering are plotted in arbitrary units. \newline

\begin{figure}[htb]
\centerline{\includegraphics[width=0.4\linewidth]{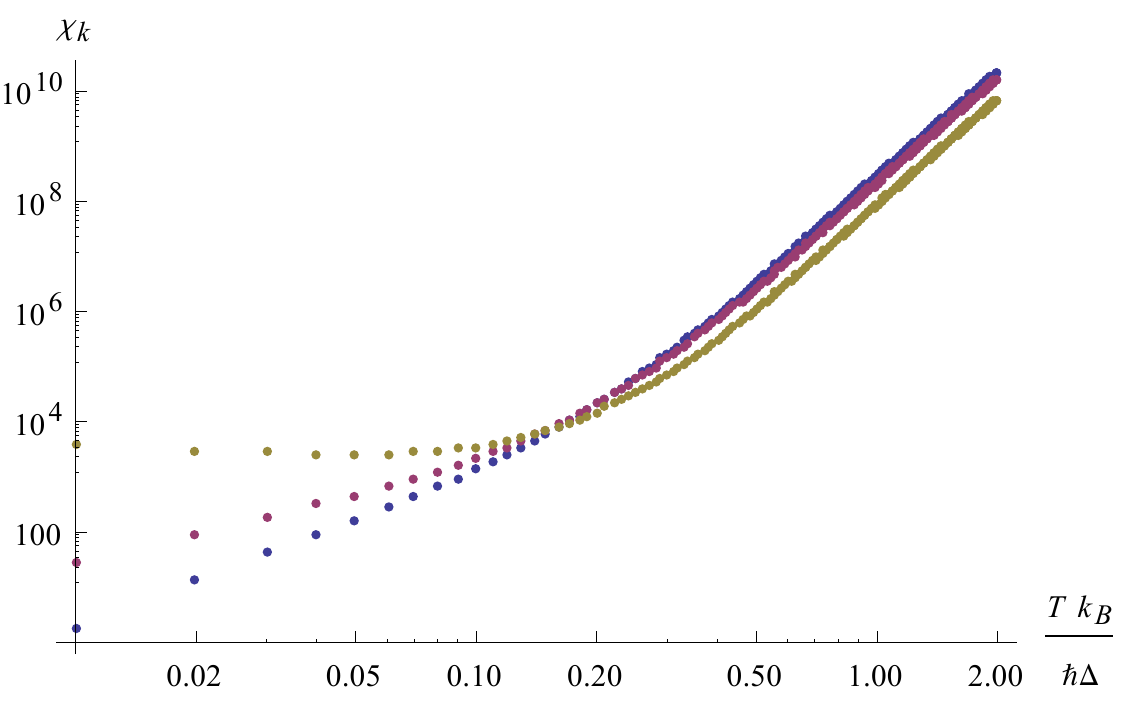}
}
\caption{Temperature dependencies of the thermal conductivity (in arbitrary units) due to the phonon-kinks scattering as function of temperature in units of characteristic kink oscillation frequency $\hbar\Delta/k_B \sim 1K$. The blue points represent $\beta =1$, the purple points represent $\beta =2$ and the brown yellow points represent $\beta =4$, with parameter $\beta$ taken from the random matrix theory \cite{Rab04}}
\label{kinkplot}
\end{figure}

\section{Conclusions}
In the forthcoming detailed publication the above theoretical results will be compared in detail with the vast available experimental data, see \cite{Mez79, Mez84} and references there in. Here we merely mention the major result of the present work. Namely, topological defects created on the dislocation lines in the form of kinks prove to be responsible for the dominant contribution to the inelastic phonon scattering at low temperatures, where the phonon-phonon scattering is vanishing. The reason for this is that unlike the thermal phonons, the kinks on the dislocation line in a crystal form non-vanishing density of mobile defects in the $T\to 0$ limit, making vibration spectrum dependent on the sample cooling history. The history is 'inscribed' by the random pattern of the local frozen irregular stresses in the crystal. This glassy state behaviour becomes a measurable 'historic imprint' when the density of kinks overcomes some characteristic value. Namely, for dominated by the phonon-kink scattering heat flow we find, that the theory predicts comparable with the experiments thermal conductivities when the density of kinks in a crystal is enough to obey the following estimate: $n_kn_d\tau_k \geq 3 \cdot 10^{11}\frac{\text{s}}{\text{m}^3}$. 

\section*{Acknowledgements}
The authors are grateful to Prof. L.P. Mezhov-Deglin for valuable discussions of the experimental data in low temperature transport anomalies in deformed crystal samples at low temperatures. S.I.M. is grateful to Prof. Beenakker and colleagues at the Lorentz  Institute for useful discussions and hospitality during his stay in Leiden in the course of this work. 

\end{widetext}

\end{document}